\documentclass{PoS}

\usepackage{siunitx} 
\usepackage{amsmath}
\usepackage{hepnames}
\usepackage{sidecap}

\title{Next-to-Leading Order QCD Corrections to\\
          Inclusive Heavy-Flavor Production in\\
          Polarized Deep-Inelastic Scattering}

\ShortTitle{NLO Inclusive HQ Production in Polarized DIS}

\author{\speaker{F. Hekhorn} and M. Stratmann\\
        Institute for Theoretical Physics, University of T\"ubingen, Auf der Morgenstelle 14,\\ 72076 T\"ubingen, Germany\\
        E-mail: \email{felix.hekhorn@uni-tuebingen.de}, \email{marco.stratmann@uni-tuebingen.de}}

\abstract{We report on a recently completed, first calculation of the full next-to-leading 
order QCD corrections for heavy flavor contributions to the inclusive structure function 
$g_1$ in longitudinally polarized deep-inelastic scattering. 
All results are derived with largely analytical methods and retain the full dependence
on the heavy quark's mass.
As a first phenomenological application, inclusive charm production 
at a future electron-ion collider and its sensitivity to the polarized gluon distribution is studied.
Theoretical uncertainties due to the residual dependence 
on unphysical factorization and renormalization scales are estimated.}

\FullConference{XXVI International Workshop on Deep-Inelastic Scattering and Related Subjects (DIS2018)\\
		16-20 April 2018\\
		Kobe, Japan}

\begin{document}

Heavy quarks (HQ) are an important and versatile laboratory for 
probing different aspects of Quantum Chromodynamics (QCD). 
Phenomenological studies of the nucleon structure in terms of parton distribution functions (PDFs)
greatly benefit from data on HQ production \cite{Butterworth:2015oua} due to the dominance of 
gluon-induced production processes already at the lowest order (LO) approximation of QCD.
In case of unpolarized deep-inelastic scattering (DIS), the charm contribution to the
structure function $F_2$ amounts to about $25\%$ \cite{Abramowicz:1900rp}
and, hence, utilizing at least the full next-to-leading order (NLO) QCD corrections \cite{Laenen:1992zk}
in quantitative analyses is a must.

Correspondingly, heavy flavor production in DIS with longitudinally polarized beams and targets,
i.e., the HQ contribution to the relevant structure function $g_1$, is expected to
reveal novel insights into the so far elusive and poorly constrained gluon helicity distribution $\Delta \Pg$.
Current uncertainties in $\Delta \Pg$ \cite{deFlorian:2008mr} 
prevent one from answering one of the most topical questions 
in Nuclear Physics, namely what is the net contribution of gluons to the spin of the proton, i.e.,
what is the value of the first moment $\int_0^1 \Delta \Pg(x,Q^2)\,dx$.
Recent data from polarized proton-proton collisions at BNL-RHIC \cite{Aschenauer:2015eha}
have revealed first evidence for a sizable contribution to the integral at medium-to-large
values of $x$ \cite{deFlorian:2014yva}, but nothing can be said about $\Delta \Pg(x,Q^2)$
for $x$ values smaller than about $0.01$.
In particular, a high-luminosity Electron-Ion Collider (EIC) \cite{Boer:2011fh}, whose  
physics case and technical realization is currently under scrutiny in the U.S.,
would uniquely offer access to a broad kinematic regime of small-to-medium momentum fractions $x$ 
in a range of virtualities $Q^2$ of the exchanged photon in DIS. 
At an EIC the charm contributions to $g_1$ in polarized DIS 
could be sizable and, hence, experimentally accessible within meaningful uncertainties.

Here, we report on a first computation of the relevant
NLO QCD corrections for HQ production in polarized DIS \cite{ref:paper} which are
mandatory to perform a meaningful and reliable phenomenological analysis of future EIC data.
Our calculation completes the existing suite of NLO results for HQ photo- and hadroproduction 
in collisions of longitudinally polarized beams and targets \cite{ref:other-nlo}.
It closely follows the technical steps outlined and used in Ref.~\cite{Laenen:1992zk} 
to derive the corresponding unpolarized results. 
In particular, largely analytical methods are adopted throughout,  
and the full dependence on the HQ's mass $m$ is retained in the final expressions
along with the other energy scale in DIS provided by the virtuality $Q$ of the photon
exchanged between the lepton and the nucleon.
$m$ acts as a natural regulator in perturbative calculations but significantly 
complicates, for instance, analytical phase-space integrations for which
an extensive list can be found in \cite{Beenakker:1988bq,Laenen:1992zk}.

\begin{figure}[t]
\begin{center}
\includegraphics[width=.95\textwidth]{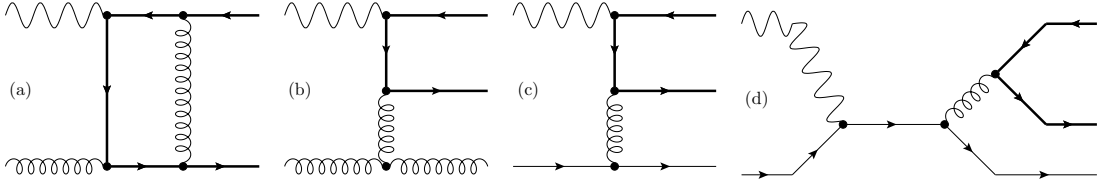}
\end{center}
\vspace{-1.0em}
\caption{Representative Feynman diagrams for different NLO contributions: 
\textbf{(a)} virtual and \textbf{(b)} real gluon emission corrections to PGF
and light-quark induced \textbf{(c)} Bethe-Heitler and \textbf{(d)} 
Compton processes.}
\label{fig:nlo}
\end{figure}

To derive the full NLO corrections to the photon-gluon fusion (PGF) process 
solely contributing to  HQ electroproduction at the Born approximation, one
has to compute several types of Feynman diagrams such as the ones illustrated in Fig.~\ref{fig:nlo}.
The regularization of intermediate singularities is performed in dimensional regularization,
with particular care for $\gamma_5$ and the Levi-Civita tensor used to project onto definite
helicity states of the initial-state photon and parton. For details, including also the renormalization of the
HQ mass and the strong coupling $\alpha_s$ as well as the mass factorization procedure, see \cite{ref:paper}
and references therein.

As in \cite{Laenen:1992zk}, our calculation \cite{ref:paper} provides results 
on single-inclusive heavy quark (or antiquark) distributions in longitudinally polarized DIS
but focuses mainly on the phenomenologically most relevant, fully inclusive HQ
contributions to the helicity-dependent structure function $g_1(x,Q^2)$. The latter results
are obtained by integration over the entire partonic phase space available and are expressed in
terms of LO and NLO PGF and genuine NLO, light-quark initiated Bethe-Heitler and Compton scaling functions 
$c_{P,\Pg}^{(0)}$, $c_{P,\Pg}^{(1)}$  and 
$c_{P,\Pq}^{(1)}$ and $d_{P,\Pq}^{(1)}$, respectively.
Due to the two-scale nature of HQ electroproduction, the scaling functions depend on 
the variables $\eta = (s-4m^2)/(4m^2)$ and $\xi = Q^2/m^2$, where $s$ denotes the partonic center-of-mass
system energy squared. 

More specifically, the charm contribution $g_1^{\Pqc}(x,Q^2,m^2)$ to longitudinally polarized DIS at NLO
accuracy can be written as a convolution of the partonic scaling functions with appropriate combinations
of helicity PDFs using $z_{\max}=Q^2/(4m^2+Q^2)$ and $\eta=\frac{1-z}{z}\frac{Q^2}{4m^2}-1$:
\begin{eqnarray}
\nonumber
&&2\,x\,g^{\Pqc}_{1}(x,Q^2,m^2) = \frac{\alpha_s(\mu_R^2)}{4\,\pi^2} \frac{Q^2}{m^2} 
\int\limits_x^{z_{max}}\frac{dz}{z} \Bigg\{ \Delta \Pg\left(\frac{x}{z},\mu_F^2\right) e_{\Pqc}^2 \,
c^{(0)}_{P,\Pg}(\eta,\xi) \\
\nonumber
&&+ 4\pi\alpha_s(\mu_R^2) \Bigg( \,\Delta \Pg\left(\frac{x}{z},\mu_F^2\right) e_c^2
\left[ c_{P,\Pg}^{(1)}(\eta,\xi) + \bar c_{P,\Pg}^{F,(1)}(\eta,\xi) \ln\left(\frac{\mu_F^2}{m^2}\right)
+ \bar c_{P,\Pg}^{R,(1)}(\eta,\xi) \ln\left(\frac{\mu_R^2}{m^2}\right)\right] \\
&&+ \!\!\!\!\sum_{\Pq=\Pqu,\Pqd,\Pqs} \left[\Delta \Pq+\Delta\Paq\right]\left(\frac{x}{z},\mu_F^2\right) \left[ e_{\Pqc}^2
\left( c_{P,\Pq}^{(1)}(\eta,\xi) + \ln\left(\frac{\mu_F^2}{m^2}\right)\bar c_{P,\Pq}^{F,(1)}(\eta,\xi)\right)
+ e_q^2\,  d_{P,\Pq}(\eta,\xi) \right] \Bigg) \Bigg\}.
\label{eq:g1contr}
\end{eqnarray}
Here, the indices $R$ and $F$ denote the contributions to the scaling functions originating from
the renormalization and factorization procedures performed at scales $\mu_R$ and $\mu_F$, respectively.
An equation similar to (\ref{eq:g1contr}) holds for the unpolarized DIS structure functions
$F_{1,2,L}^{\Pqc}(x,Q^2)$ when both the PDFs and the scaling functions are substituted
appropriately, see \cite{ref:paper}. We note that in \cite{ref:paper} also the unpolarized scaling functions
have been re-derived as a benchmark calculation and successfully compared to the results given in \cite{Laenen:1992zk}.
In addition, various important limits of the polarized and unpolarized scaling functions
[threshold ($\eta\to 0$), high-energy ($\eta\to\infty$), photoproduction ($Q^2\to 0$)] have been thoroughly studied in
\cite{ref:paper} and compared to results available in the literature \cite{ref:compare}.

\begin{figure}[t]
\begin{center}
\includegraphics[width=.9\textwidth]{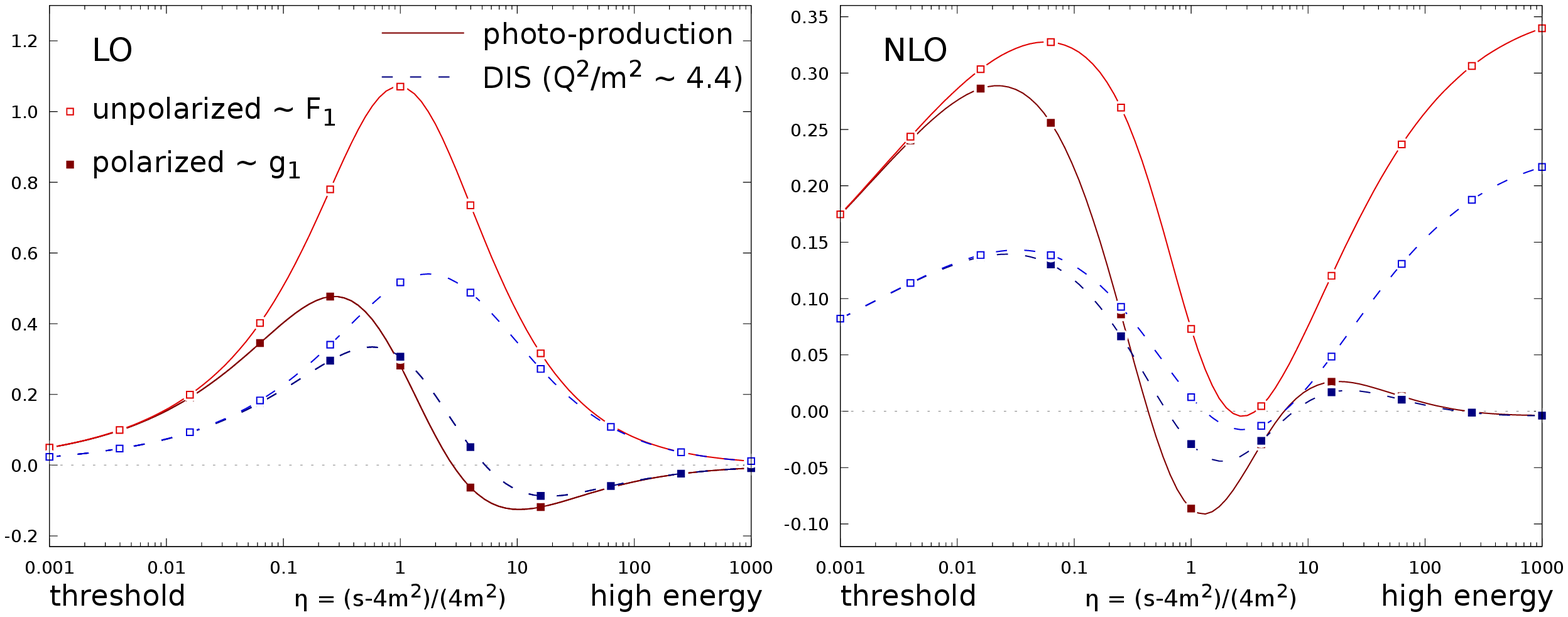}
\end{center}
\vspace{-1.0em}
\caption{LO and NLO polarized (full symbols) and unpolarized (open symbols) PGF scaling functions 
for two values of $\xi=Q^2/m^2$, one the DIS (dashed lines) and one close to the photoproduction (solid lines) regime.}
\label{fig:cgTP-new}
%
\begin{center}
\includegraphics[width=.9\textwidth]{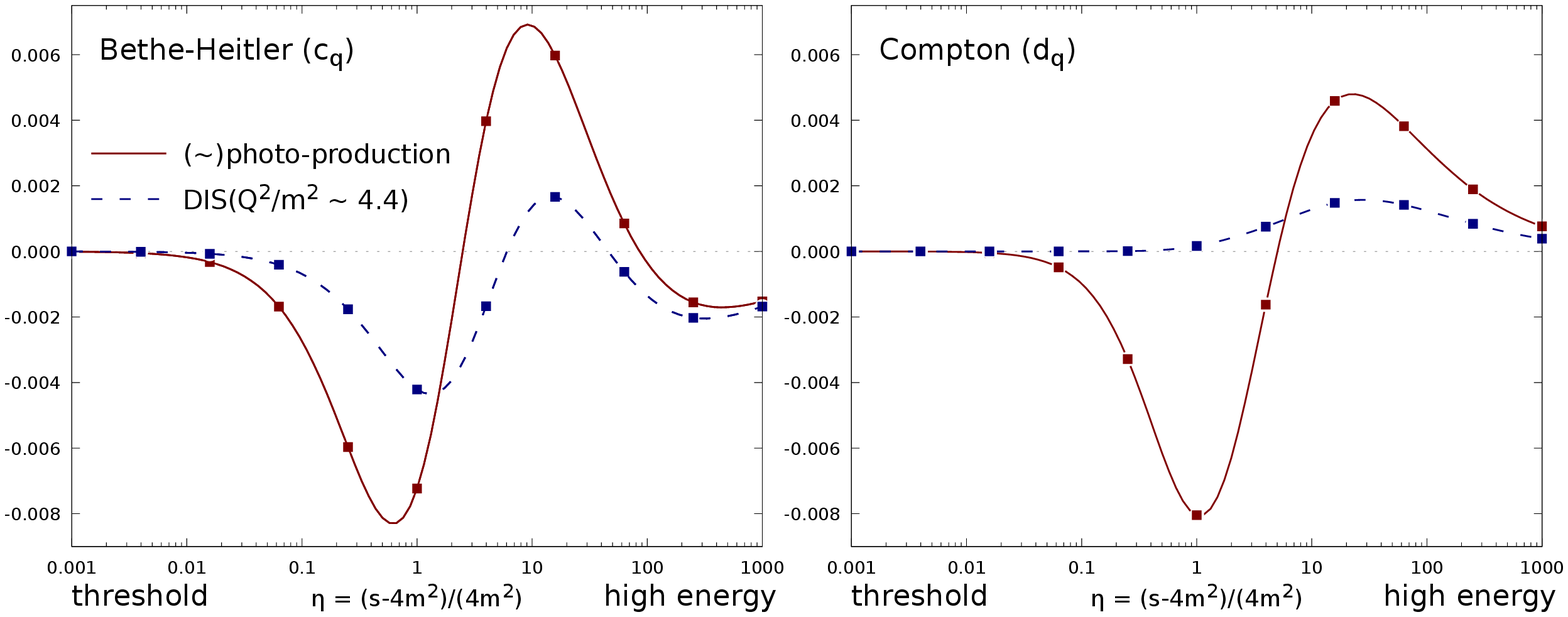}
\end{center}
\vspace{-1.0em}
\caption{As in Fig.~\ref{fig:cgTP-new} but now for the NLO polarized 
quark scaling functions $c_{P,\Pq}^{(1)}$ and $d_{P,\Pq}^{(1)}$.} 
\label{fig:cdqP-new}
\end{figure}

The LO and NLO PGF coefficient functions $c_{k,\Pg}^{(0)}$ and $c_{k,\Pg}^{(1)}$ are shown in Fig.~\ref{fig:cgTP-new}
for $k=P$ and $T$ relevant for the computation of $g_1$ and $F_1$, respectively. 
As is expected, close to threshold, $s\to 4m^2$, the polarized and unpolarized results approach each other whereas
in the high-energy limit, $s\to \infty$, $c_{P,\Pg}$ vanishes but $c_{T,\Pg}$ reaches a plateau value. In between, the
dependence on $\eta$ (and $\xi$, see \cite{ref:paper}) is in general rather non-trivial.
The genuine NLO polarized light-quark scaling functions $c_{P,\Pq}^{(1)}$ and $d_{P,\Pq}^{(1)}$ are depicted in 
Fig.~\ref{fig:cdqP-new}. Due to Furry's theorem there is no
interference term between the Bethe-Heitler [Fig.~\ref{fig:nlo}~\textbf{(c)}]
and the Compton process [Fig.~\ref{fig:nlo}~\textbf{(d)}], 
which would give rise to a contribution $\sim e_c e_q$ in the electrical quark charges in Eq.~(\ref{eq:g1contr}).
Also note that the Compton process contains logs of the form $\ln(Q^2/m^2)$ such that $d_{P,\Pq}^{(1)}$
does not have a finite photoproduction limit $Q^2\to 0$.

\begin{figure}[t]
\begin{center}
\includegraphics[width=.8\textwidth]{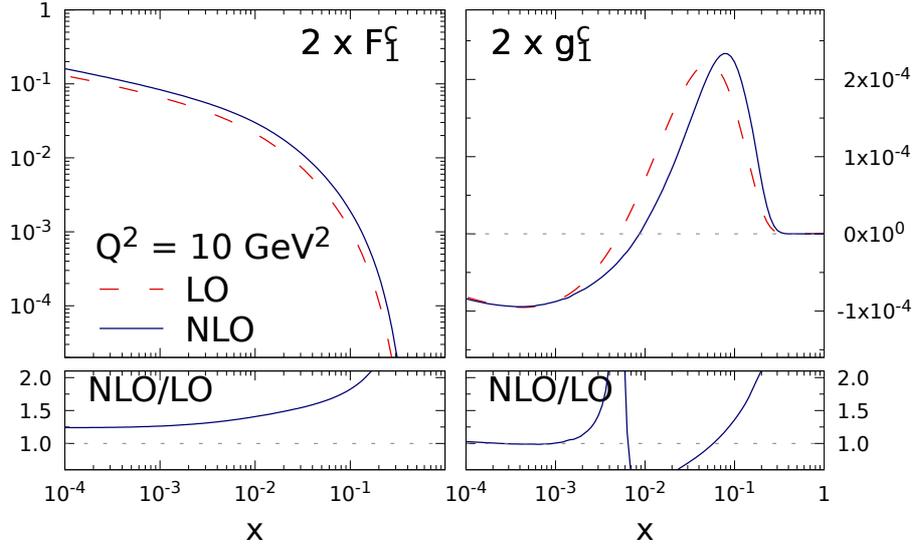}
\end{center}
\vspace{-1.0em}
\caption{The DIS charm structure functions $2\,x\,F_1^{\Pqc}$ (left-hand-side) and $2\,x\,g_1^{\Pqc}$ (right-hand-side) 
at LO and NLO accuracy as a function of x for $Q^2 = \SI{10}{\GeV^2}$. The lower panels
show the respective ``K-factor'', see text. All results were
obtained using $m = m_{\Pqc} = \SI{1.5}{\GeV}$ and $\mu_F^2 = \mu_R^2 = 4m^2 + Q^2$.}
\label{fig:F1g1}
\end{figure}

In Fig.~\ref{fig:F1g1} we compare the DIS charm structure functions $F_1^{\Pqc}$ and $g_1^{\Pqc}$ 
as a function of $x$ for $Q^2=\SI{10}{\GeV^2}$. In the calculation we have used the
MSTW2008 \cite{Martin:2009iq} and DSSV2014 \cite{deFlorian:2014yva} sets of unpolarized and helicity-dependent PDFs, respectively,
and our default choice of scales: $m = m_{\Pqc} = \SI{1.5}{\GeV}$, $\mu_F^2 = \mu_R^2 = \mu_0^2= 4m^2 + Q^2$.
The lower panels show the respective ``K-factor'' as a measure of the relevance of the NLO corrections.
As can be seen, both the polarized and unpolarized structure functions do receive 
sizable corrections at NLO that are non-uniform in the kinematic variables $x$ and $Q^2$.
As $g_1^{\Pqc}$ refers to a cross section difference, it is not a strictly positive number and, hence,
the visualization of the K-factor is plagued by the different nodes in the LO and NLO results with respect to $x$.

\begin{SCfigure}
\includegraphics[width=.4\textwidth]{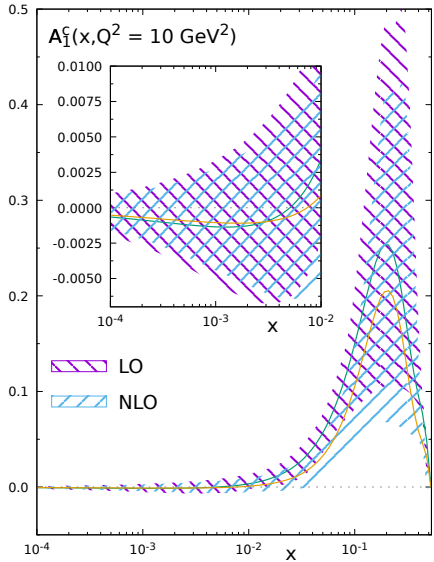}
\caption{The double spin asymmetry $A_1^{\Pqc}$
for charm electroproduction at LO and NLO accuracy for $Q^2 = \SI{10}{\GeV^2}$. 
The shaded bands illustrate the allowed spread in helicity PDFs as estimated by the DSSV group.
The inset zooms into the phenomenologically interesting small-$x$ region.}
\label{fig:A1-pdf}
\end{SCfigure}

The experimentally relevant double-spin asymmetry, defined as $A_1^{\Pqc}(x,Q^2) \equiv g_1^{\Pqc}(x,Q^2)/F_1^{\Pqc}(x,Q^2)$, 
is shown in Fig.~\ref{fig:A1-pdf} at $Q^2=\SI{10}{\GeV^2}$
along with an uncertainty band obtained from currently allowed variations of
helicity PDFs as estimated by the DSSV group \cite{deFlorian:2014yva}.
It turns out that even the sign of $g_1^{\Pqc}$ is presently unknown in the small-$x$ region, 
which can be entirely traced back to the poorly constrained $\Delta g$ in that kinematic regime.
Also the relative importance of the PGF process, which strongly dominates in $F_1^{\Pqc}$,
with respect to genuine NLO light-quark initiated contributions is largely uncertain. For instance, for
the default set of DSSV both gluons and light quarks contribute roughly on equal footing to $g_1^{\Pqc}$.
Clearly, a sufficiently good measurement of $A_1^{\Pqc}$ at an EIC has the potential to reduce our current
ignorance of $\Delta \Pg$ at small momentum fractions. In order to do so, possible future EIC data
on $A_1^{\Pqc}$ at $x\simeq 10^{-3}$ and $Q^2=\SI{10}{\GeV^2}$ have to achieve 
at least a precision at the level of $\mathcal O(10^{-3})$
as can be estimated from the spread of $A_1^{\Pqc}$ shown in the inset in Fig.~\ref{fig:A1-pdf}.

\begin{figure}[t]
\begin{center}
\includegraphics[width=.9\textwidth]{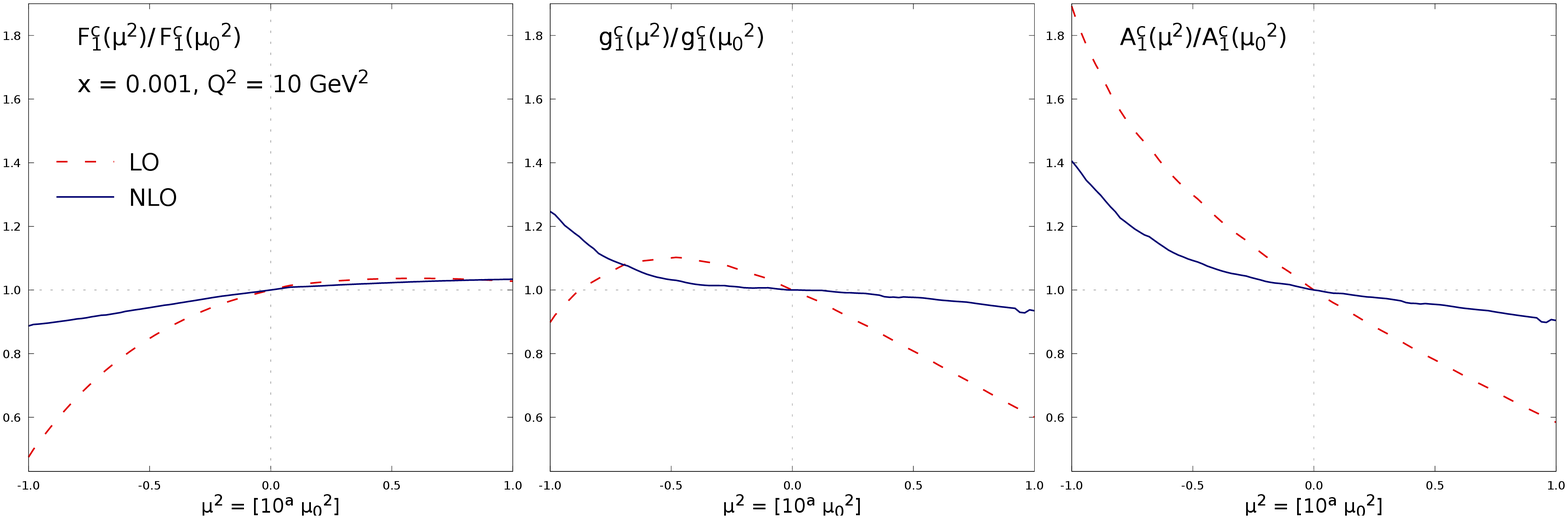}
\end{center}
\vspace{-1.0em}
\caption{Scale dependence of $F_1^{\Pqc}$, $g_1^{\Pqc}$, and $A_1^{\Pqc}$ 
for $\mu^2=\mu_F^2 = \mu_R^2 = 10^a \mu_0^2$ with $\mu_0^2 = 4m^2+Q^2$.}
\label{fig:F1g1A1-mu2}
\end{figure}

Finally, in Fig.~\ref{fig:F1g1A1-mu2} we investigate the dependence of $F_1^{\Pqc}$, $g_1^{\Pqc}$, and $A_1^{\Pqc}$ on
common variations of the unphysical factorization and renormalization scales by a factor of ten around
our default choice $\mu^2_F=\mu_R^2=\mu_0^2=4m^2+Q^2$. As expected, the NLO results exhibit a weaker dependence
on the actual choice of $\mu^2_0$. In general, the scale dependence depends on the chosen kinematics, see Ref.~\cite{ref:paper}, where
also independent variation of $\mu_F$ and $\mu_R$ have been studied.


%

\end{document}